\documentclass[aps,prapplied,reprint,nofootinbib,
superscriptaddress,
amsmath,amssymb,
]{revtex4-2}

\usepackage{graphicx}
\usepackage{dcolumn}
\usepackage{bm}
\usepackage{physics}
\usepackage{braket}
\usepackage{xcolor}
\usepackage{nicefrac}
\usepackage{upgreek}
\usepackage[colorlinks=true, linkcolor=blue, citecolor=blue, urlcolor=blue]{hyperref}

\begin{document}


\title{Magnetic Communication with an Acoustically Actuated Magnetoelectric Resonator and a Quantum Diamond Magnetometer}


\author{C. T.-K. Lew}
\email{christopher.tao-kuan.lew@rmit.edu.au}
\thanks{These authors contributed equally to this work.}
\affiliation{
Department of Physics, School of Science, RMIT University, VIC, 3001, Australia
}

\author{A. D. M. Charles}
\email{andrew.charles1@defence.gov.au}
\thanks{These authors contributed equally to this work.}
\affiliation{
Platforms Division, Defence Science and Technology Group, 506 Lorimer Street, Fishermans Bend, VIC 3207, Australia
}

\author{B. C. Gibson}
\affiliation{
Department of Physics, School of Science, RMIT University, VIC, 3001, Australia
}

\author{J.-P. Tetienne}
\affiliation{
Department of Physics, School of Science, RMIT University, VIC, 3001, Australia
}

\author{D. A. Broadway}
\email{david.broadway@rmit.edu.au}
\affiliation{
Department of Physics, School of Science, RMIT University, VIC, 3001, Australia
}

\date{\today}

\begin{abstract}
Wireless communication via propagating magnetic fields is a communication modality that has recently garnered significant interest for short-to-medium range communication in conductive mediums, such as underwater and underground, where existing approaches utilizing electric fields are highly inefficient. Typical implementations of magnetic communication make use of loop antennas as both the transmitter and receiver, with the sensitivity and frequency response scaling with and inversely with the loop cross-sectional area, respectively. Here, we explore an alternative hybrid magnetic communication system consisting of an highly radiation efficient and compact acoustically actuated magnetoelectric resonator as the transmitter, and a highly sensitive micrometer scale quantum magnetometer based on nitrogen-vacancy centers in diamond as the receiver, with their core properties unconstrained by size. We demonstrate amplitude and phase-encoded transmission and reception of AC magnetic fields at $f_{\mathrm{AC}} = 20$ kHz, achieving a sensitivity of 50 $\mathrm{pT/\sqrt{Hz}}$ and 1.2 $\mathrm{mrad/\sqrt{Hz}}$, respectively. This work establishes the use of hybrid magnetoelectric resonator and  quantum diamond magnetometer communication system as a viable alternative to existing loop-based approaches.
\end{abstract}

\maketitle



\section{Introduction}
The ability to generate, transmit, and receive electromagnetic (EM) signals is integral to the way information is transferred wirelessly \cite{Goldsmith2005}. In practical scenarios outside ideal laboratory conditions, the propagating medium can play a significant role in distorting, attenuating, or even fully suppressing signals sent between the transmitter and receiver \cite{Saunders2024}. Specifically, high frequency EM fields are highly attenuated in conductive mediums such as metals, geological materials (i.e., rocks, soils, etc.), and seawater. The attenuation factor is characterized by the skin depth, $\delta = \sqrt{1/\pi \upmu_{r} \sigma f_{\mathrm{AC}}}$, where $f_{\mathrm{AC}}$ is the signal frequency, $\upmu_{r}$ and $\sigma_c$ are the relative magnetic permeability and electrical conductivity of the material, respectively. Naturally, the use of very low frequency ($3-30$ kHz) signals is favorable in highly conductive mediums to minimize the skin depth. However, coupling sources and detectors to the electric field component of the transmitting EM field remains a challenge. As such, low frequency magnetic fields for short-to-medium range communication in conductive mediums is seen as a promising solution.

Traditionally, magnetic communication rely on the use of loop antennas as both the transmitter and receiver due to their simplicity and low noise performance \cite{Tumanski2007}. On the transmitter side, loop antennas in the near-field region (d $<\lambda/2\pi$) are highly inefficient due to their low source impedance \cite{Satitchantrakul2026,Keller2023}. An alternative magnetic transmitter technology that has recently garnered significant attention is the acoustically driven magnetoelectric (ME) resonator, which combines both the piezoelectric and magnetostriction effect to efficiently convert between magnetic and electric fields via strain mechanical coupling. Compared to traditional loop-based transmitter approaches, ME resonators can offer four to five orders of magnitude reduction in size \cite{Chen2020,Li2022}, in addition to improved radiation efficiency \cite{Wang2011}.



\begin{figure*}[t]
    \centering
    \includegraphics[width=6in]{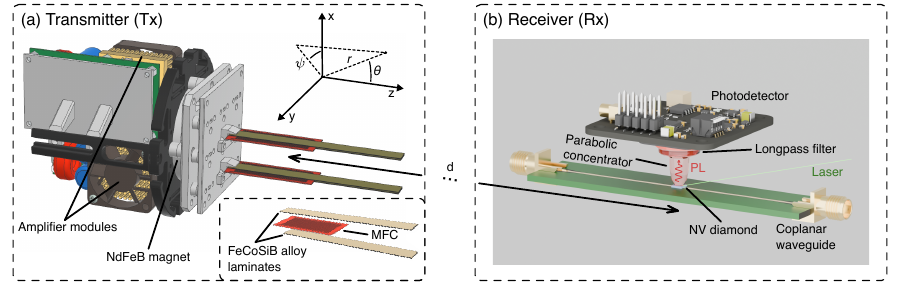}
    \caption{\textbf{Hybrid magnetic communication system.} (a) Simplified schematic of the magnetoelectric (ME) resonator transmitter subsystem and the coordinate system used, including the amplifier module, bias magnet, and end-clamped multi-layered ME resonators attached via a 3D-printed structure. An expanded view of a single resonator element with the two FeCoSiB alloy/epoxy laminates bonded symmetrically above and below the Macro-Fiber Composite (MFC) package is shown in the inset. (b) Simplified schematic of the NV diamond receiver subsystem separated by distance, d, from the ME resonator. At the core of the receiver sensing element is the NV diamond, which is placed on a coplanar waveguide for coherent microwave (MW) excitation. The photoluminescence (PL) from the optically excited NV ensemble is collected using a parabolic concentrator, filtered with a longpass filter, and measured using a photodetector.} 
    \label{F1}
\end{figure*}

On the receiver side, a loop antenna with high sensitivity can be achieved at the expense of a large loop cross-sectional area. For example in Ref.~\cite{Cohen2009}, a sensitivity of ${\sim}2 \; \mathrm{fT/\sqrt{Hz}}$ is achieved at $f_{\mathrm{AC}} = 20$ kHz using an antenna with a 1.69~m$^2$ loop cross-sectional area. Quantum sensors are an emerging type of magnetometer technology that can overcome classical limits and operate as highly sensitive and ultracompact receivers. In particular, nitrogen-vacancy (NV) centres in diamond have been demonstrated to operate at sub-$\mathrm{pT/\sqrt{Hz}}$ sensitivities with the sensing element on the micrometer to millimeter length scale \cite{Barry2024,Wolf2015}. Furthermore, the NV frequency response can be dynamically tunable from DC to GHz \cite{Degen2017,Pham2012}, enabling broadband coverage. Despite these appealing properties, the use of NV diamond as receivers currently remains largely unexplored, with near DC ($< 500$ Hz) \cite{Zhu2024,Krumins2025} and GHz \cite{Shao2016,Zeng2025} communication only been recently demonstrated.

In this work, we demonstrate a compact magnetic communication system operating at $f_{\mathrm{AC}}=20$ kHz utilizing a ME resonator as the transmitter and a NV magnetometer as the receiver. Both amplitude and phase-encoded transmission and reception were explored, with reception of phase-modulated information achieving a higher signal-to-noise ratio (SNR). The amplitude and phase sensitivity of the NV diamond receiver was determined to be 50 $\mathrm{pT/\sqrt{Hz}}$ and 1.2 $\mathrm{mrad/\sqrt{Hz}}$ at $B_{\mathrm{AC}}=34.3$~nT, respectively, predominately limited by residual laser-intensity noise. Lastly, we consider operation in relevant conductive mediums and estimate transmission and reception distances achievable. This work expands the applicability of alternative magnetic communication approaches using ME resonators and quantum magnetometers. 

\section{Experiment}

\subsection{ME resonator transmitter}
The ME resonator transmitter is a multi-layered heterostructure design originally developed by Wang \cite{Wang2011,WANG2012} and Gao et al. \cite{GAO2011} and consists of a flexible piezoelectric patch sandwiched symmetrically by polymer bonded amorphous metal alloy magnetostrictive laminates (see Fig.~\hyperref[F1]{\ref{F1}(a)} inset). The piezoelectric patch is formed from a micro-structured layout of aligned piezoceramic fibers and interdigitated electrodes known as a Macro Fiber Composite (MFC). Two sets of magnetostrictive laminates were fabricated from a FeCoSiB amorphous metal alloy, with each laminate consisting of 9 layers bonded together with an epoxy resin.

Two transmitter units were manufactured and end-clamped to a 3D-printed structure, with a NdFeB magnet positioned ~6 mm behind the transmitter units in order to bias and align the magnetic domains of the amorphous alloy layers, maximizing the ME effect (Fig.~\hyperref[F1]{\ref{F1}(a)}). For all measurements presented in this work, only the top transmitter unit was used. A $\pm$200~V linear amplifier module was used to drive the ME resonator. The amplifier's 20 V$_{\mathrm{drive}}$/V$_{\mathrm{in}}$ gain restricts the input voltage range to V$_{\mathrm{in}}=0-10$ V, delivering a maximum driving output of V$_{\mathrm{drive}}=200$ V. However, the ME resonator design can withstand applied driving voltages between $-500$ V and 1500 V.


\subsection{NV diamond receiver}
Measurements were performed using a modified version of the experimental setup described in Lew et al. \cite{Lew2026} using a 0.5 mm thick NV-doped diamond sample (Element Six, DNV$-$B1), with a typical NV density of 300 ppb. 532 nm laser pulses controlled using an acousto-optic modulator are used to initialize and optically spin polarize ($t_{I}=15\; \upmu$s) the NV ensemble to the $m_s=0$ ground state and readout ($t_{R}=30\; \upmu$s) the final spin state population. The photoluminescence (PL) from the optically excited NV ensemble is collected using a compound parabolic concentrator, filtered with a $\lambda=655$~nm longpass filter, and measured using a large area photodetector (Fig.~\hyperref[F1]{\ref{F1}(b)}). 

The laser power at the diamond sample is approximately 1.5 W, resulting in around 11.9 mW of PL collected and measured. A small portion of the laser power is picked off and measured using a second reference large area photodetector for common-mode rejection. The differential photovoltage between the two photodetectors is digitized using a data acquisition unit. A static bias magnetic field, $B_0$, is applied along the $\braket{111}$ direction to lift the degeneracy between the $m_s =\pm1$ states of the NV electronic spin. Coherent microwave (MW) pulses were delivered to the sample using a 1.2 mm wide central stripline coplanar waveguide. The NV diamond receiver is aligned approximately in the same plane as the ME resonator along the z-axis in the laboratory frame (see Fig.~\hyperref[F1]{\ref{F1}(a)}) where it is expected the magnetic field generated by the ME resonator is maximized (see Appendix \ref{ME_pattern} for expected radiation pattern).

\section{Results}

\subsection{AC magnetometry method}
\begin{figure}[t]
    \centering    \includegraphics[width=3.375in]{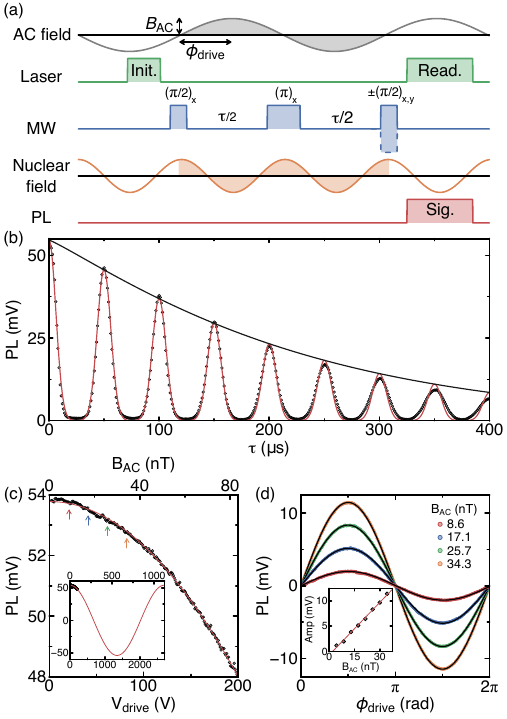}
    \caption{\textbf{NV diamond ensemble AC magnetometry.} (a) AC magnetic field measurement scheme based on the Hahn-echo pulse sequence. (b) Example spin-echo measurement. Collapses and revivals are due to interactions with the $^{13}$C nuclear spin bath. The black solid line is the exponential decay fit to the spin echo envelop, and the red solid line is the overall fit to the observed spin-echo modulation. (c) Measured spin-echo signal as a function of V$_{\mathrm{drive}}$. For $f_{\mathrm{AC}} = 20$~kHz ($\tau = 50 \; \upmu$s), the NV spin ensemble accumulates one phase cycle (red solid line) every $B_{\mathrm{AC}} = 1.12 \; \upmu$T (inset), which is used to determine the conversion factor between V$_{\mathrm{drive}}$ and $B_{\mathrm{AC}}$. (d) Measured PL response as a function of the transmitted AC magnetic field phase for different magnetic field strengths (colored arrows in (c)). Black solid lines are sinusoidal fits to the experimental data. The extracted amplitude of the sinusoid is plotted as a function of magnetic field strength in the inset, revealing a linear dependence.} 
    \label{F2}
\end{figure}

For magnetic field sensing in the kHz regime, the canonical quantum sensing approach is performed by applying a Hahn-echo control sequence depicted in Fig.~\hyperref[F2]{\ref{F2}(a)}. The NV ensemble spin coherence will accumulate a net phase in the presence of an oscillating AC magnetic field. When the sensing interval, $\tau$, matches the period and the MW control pulses coincide with the nodes of the AC magnetic field, the accumulated phase is maximized. The acquired phase, $\delta\phi = 4 \gamma_{\mathrm{NV}} B_{\mathrm{AC}}\tau$, scales with the AC magnetic field strength, $B_{\mathrm{AC}}$, where $\gamma_{\mathrm{NV}} =28$ MHz/mT is the NV electron spin gyromagnetic ratio \cite{Maze2008,Taylor2008}.

An additional oscillating AC magnetic field source present inherent to the diamond arises from the $1.1\%$ naturally abundant $^{13}$C nuclear spin bath. The effect of this nuclear spin bath is observed in the collapse and revival of the spin-echo signal in Fig.~\hyperref[F2]{\ref{F2}(b)} (in the absence of an external AC magnetic field) as the $^{13}$C nuclear spin bath precesses at a Larmor frequency $\omega_L = \gamma_{^{13}C} B_0$,  where $\gamma_{^{13}C} = 10.708$ MHz/T is the carbon gyromagnetic ratio. The rate of collapse and revival occurs at half the $^{13}$C Larmor frequency, which can be tuned via the strength of the static magnetic field applied. For Fig.~\hyperref[F2]{\ref{F2}(b)}, a static magnetic field of $B_0 =3.74$ mT is applied along the NV axis parallel to the $\braket{111}$ direction, coinciding with the first revival at $\tau=50\;\upmu$s. Sensitivity to external AC magnetic fields at $f_{\mathrm{AC}} = 20$ kHz is therefore maximized and contribution from $^{13}$C is effectively suppressed.

The spin coherence time, $T_{2}$, can be extracted from the spin-echo signal in Fig.~\hyperref[F2]{\ref{F2}(b)}, where the experimental data is fitted with the following expression \cite{Lin2021}

\begin{equation}
\mathrm{PL} = A e^{-(\tau/T_{2})^{n}} \sum_{i}\exp[-(\tau-iT_{\mathrm{rev.}})^2/T_{\mathrm{dec.}}^{2}] \: .
\label{eq:eq1}
\end{equation}

\noindent{$A$} is the signal amplitude, $n$ the stretched exponential factor, $T_{\mathrm{rev.}}$ is the period of revival, and $T_{\mathrm{dec.}}$ is the relaxation time of the first collapse. From the fit to the experimental data, $T_2=231 \pm 3 \; \upmu$s and $n = 1.12 \pm 0.02$, limited by dipolar interaction with the inhomogeneous substitutional nitrogen (P1-center) spin bath \cite{Park2022,Stanwix2010}.

\begin{figure}[t]
    \centering
  \includegraphics[width=3.375in]{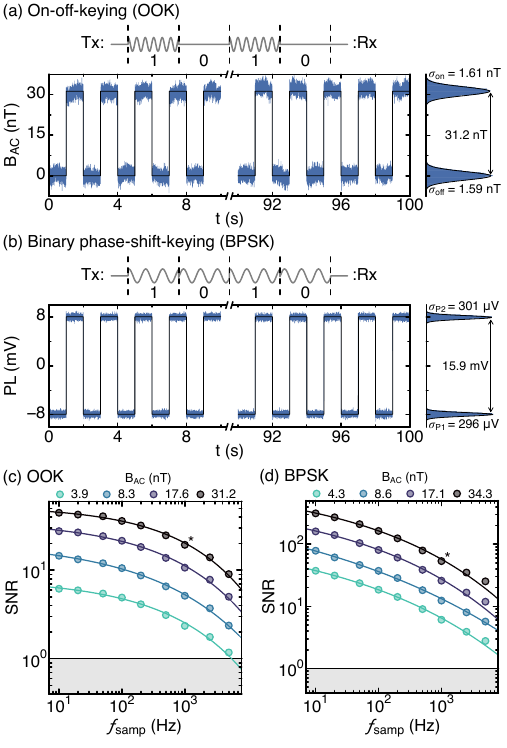}
    \caption{\textbf{Demonstration of transmission and reception of amplitude and phase modulated magnetic fields.}(a) Example 100~s time trace of the measured 1~Hz amplitude-modulated ($B_{\mathrm{AC}}=0$ and $31.2$ nT, $\phi_{\mathrm{drive}} = \pi/2$) AC magnetic field waveform (top inset) transmitted over d${\sim}0.9$~m and sampled at $f_\mathrm{samp} = 1$ kHz. The corresponding histogram is shown on the right, where the SNR is calculated from the double Gaussian fit to the experimental data. (b) Example 100~s time trace of the measured phased-modulated ($\phi_{\mathrm{drive}} = 3\pi/4$ and $5\pi/4$, $B_{\mathrm{AC}}=34.3$ nT) AC magnetic field. The transmitted phase-modulated AC magnetic field waveform is shown in the top inset and the corresponding histogram is shown on the right. (c) Measured SNR as a function of the NV magnetometer sampling rate for different magnetic field amplitudes for amplitude modulation. Solid lines are fits to the data. (d) Measured SNR as a function of the NV magnetometer sampling rate at different magnetic field amplitudes for phase modulation. Solid lines are fits to the data. The data points corresponding to (a) and (b) are denoted with an asterisk.} 
    \label{F3}
\end{figure}

The phase accumulated by the NV ensemble is observed as a spin-state population difference fluorescence signal, which is modulated by a sinusoid. Over one period of fluorescence oscillation ($\delta\phi=2\pi$) for a $f_{\mathrm{AC}} = 20$ kHz external AC magnetic field, the corresponding AC magnetic field strength is $B_{\mathrm{AC}} = 1.12 \; \upmu$T. This provides a straightforward method to calibrate the NV magnetometer response, as shown in Fig.~\hyperref[F2]{\ref{F2}(c)}. Here, the external AC magnetic field is generated by the acoustically actuated ME resonator situated d${\sim}0.2$~m away from the NV magnetometer and the sensing interval is fixed at $\tau=50\;\upmu$s. The driving voltage of the ME resonator acts as a proxy for adjusting the AC magnetic field strength. By fitting the NV magnetometer response to a sinusoid and extrapolating out to one period shown in the inset of Fig.~\hyperref[F2]{\ref{F2}(c)}, a conversion factor of 417 pT/$\mathrm{V_{drive}}$ was determined.

For external AC magnetic fields that can be phase synchronized, it is favorable to operate the NV magnetometer in phase-sensitive slope detection mode, where the NV response is most sensitive (and linear) near zero AC magnetic field strength \cite{Degen2017}. This is achieved by adjusting the final $(\pi/2)_y$ readout pulse to be orthogonal to the first $(\pi/2)_x$ pulse. Note the results presented previously in Fig.~\hyperref[F2]{\ref{F2}(b)}$-$\hyperref[F2]{(c)} were performed using a final $(\pi/2)_x$ pulse (i.e., phase insensitive variance detection mode). We verify the phase response of the NV ensemble by sweeping the phase, $\phi_{\mathrm{drive}}$, from 0 to 2$\pi$ of the AC magnetic field transmitted by the ME resonator at selective magnetic field strengths in Fig.~\hyperref[F2]{\ref{F2}(d)}, revealing a sinusoidal response (black solid lines). The amplitude of the sinusoidal fit is plotted as a function of AC magnetic field strength in the inset of Fig.~\hyperref[F2]{\ref{F2}(d)}, where a linear conversion factor of 307~$\upmu$V/nT was determined. For the remainder of this work, all measurements will be performed using phase-sensitive slope detection mode (i.e., final $(\pi/2)_y$ pulse), where the NV magnetometer is phase synchronized to the external AC magnetic field transmitted by the ME resonator. 

\subsection{Magnetic communication demonstration}
We now demonstrate the NV magnetometer receiving test data bit stream between 0 and 1 by amplitude modulating the ME resonator every second (black trace in Fig.~\hyperref[F3]{\ref{F3}(a)}). In other words, the bit values are encoded using on-off-keying (OOK). The base Hahn-echo sequence in Fig.~\hyperref[F2]{\ref{F2}(a)} lasts a total of $100 \; \upmu$s and is repeated a second time, with the final $(\pi/2)_y$ pulse phased shifted by $180^\circ$ to cancel out low frequency noise arising from charge state and laser intensity fluctuations \cite{Hart2021}. As such, the maximum effective sampling rate of the NV magnetometer is $f_{\mathrm{samp}} = 1/(2 \times 100 \; \upmu \mathrm{s})=5$~kHz. Further averaging and decimation is implemented post-processing when necessary. For Fig.~\hyperref[F3]{\ref{F3}(a)}, 5 averages per data point was implemented, reducing the effective sampling rate to 1 kHz. Clear distinction between the on/off (1/0) state is observed for a $B_{\mathrm{AC}}=31.2$~nT transmitted magnetic field. 

To quantify the clarity of the signal reception, we consider the histogram distribution of magnetic field values for the 100~s time trace in Fig.~\hyperref[F3]{\ref{F3}(a)}. Assuming white noise, a SNR${\sim}19$ was extracted from a double Gaussian fit to the distribution. In units of sensitivity, this equates to 50 pT/$\sqrt{\mathrm{Hz}}$ (see Appendix \ref{sens} for more in-depth sensitivity analysis). We note a small but consistently larger spread in the field distribution measured during the on state, suggesting a slight increase in noise when the ME resonator is switched on.  The SNR for OOK reception is plotted as a function of $f_{\mathrm{samp}}$ in Fig.~\hyperref[F3]{\ref{F3}(c)} for selective AC magnetic field strengths. An empirical exponential fit to the data of the form $\mathrm{SNR} = a_1 \exp \left( -(f_{\mathrm{samp}})^{a_2}/a_3 \right)$ was found to accurately describe the observed trends, where $a_1,a_2,$ and $a_3$ are free fitting parameters. For the highest AC magnetic field strength considered ($B_{\mathrm{AC}}=31.2$ nT), the SNR asymptotes toward low sampling rates and reaches $a_1{\sim}53$. In other words, the noise profile is random in nature at shorter timescales (high sampling rate) and an increase in colored noise occurs at longer timescales (low sampling rate) such that averaging no longer improves the SNR.

The NV magnetometer is also compatible with receiving and demodulating phase-modulated encoding schemes (i.e., phase-shift-keying (PSK)). Test data bit stream between 0 and 1 is encoded in the phase of the AC magnetic field transmitted by the ME resonator every second (black trace in Fig.~\hyperref[F3]{\ref{F3}(b)}). Clear phase separability between the (0/1) state encoded as phase jumps between ($3\pi/4$ rad) and ($5\pi/4$ rad) is observed, where the NV response is approximately linear around $\phi_{\mathrm{drive}}=\pi$~rad (see Fig.~\hyperref[F2]{\ref{F2}(d)}). A higher SNR of ${\sim}53$ (extracted from the histogram distribution of measured PL values) is achieved for similar AC magnetic field strength compared to amplitude modulation. For $B_{\mathrm{AC}}=34.3$ nT, the phase sensitivity equates to 1.15 m$\mathrm{rad}/\sqrt{\mathrm{Hz}}$. The SNR for PSK reception is plotted as a function of $f_{\mathrm{samp}}$ in Fig.~\hyperref[F3]{\ref{F3}(d)} for selected AC magnetic field strengths. From the exponential empirical fit to the experimental data, the SNR asymptotes and reaches $a_1{\sim}960$ for the highest AC magnetic field strength considered ($B_{\mathrm{AC}}=34.3$ nT).


\section{Discussion and outlook}
\begin{figure}[t]
    \centering
    \includegraphics[width=3.375in]{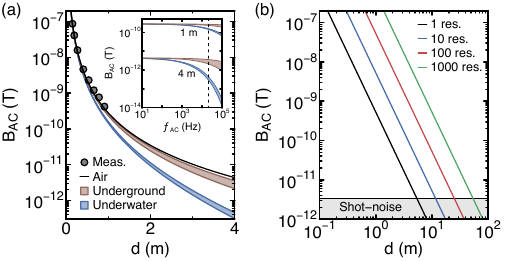}
    \caption{\textbf{Magnetic field strength at larger distances.} (a) Measured and calculated magnetic field strength drop-off as a function of transmitter-receiver distance in air (black), underground (brown), and underwater (blue), accounting for additional attenuation due to skin depth at $f_{\mathrm{AC}} = 20$ kHz. The inset plots the calculated magnetic field strength attenuation as a function of AC frequency at 1~m and 4~m. The vertical dotted line indicates $f_{\mathrm{AC}} = 20$ kHz used in this work. (b) Extrapolated magnetic field strength (in air) expected when driving multiple transmitters in parallel at full power (V$_{\mathrm{drive}} = 200$ V) as a function of transmitter-receiver distance. Horizontal line indicate current magnetic shot noise of the NV magnetometer sampling at $f_\mathrm{samp} = 100$ Hz.} 
    \label{F4}
\end{figure}

For practical communication scenarios, the distance between the transmitter and receiver will be separated meters apart. To understand the expected degraded performance of the communication system, the AC magnetic field strength with the ME resonator driven at half maximum power (V$_{\mathrm{drive}} = 100$ V) was measured as a function of distance up to ${\sim}0.9$~m (limited practically by physical space), as shown in Fig.~\hyperref[F4]{\ref{F4}(a)}. The AC magnetic field strength decays proportional to 1/d$^{3}$ as expected in the near-field regime (black solid line). In relevant environments, such as underground and underwater, additional attenuation due to the skin depth is expected, which we calculate and plot in Fig.~\hyperref[F4]{\ref{F4}(a)}. Here, the electrical conductivity of soil and seawater is estimated to be in the ranges of $0.03-0.4$ S/m \cite{Ma2015,Yan2017} and $4-5.5$ S/m \cite{Li2019,Kawamura2025}, respectively. 

Significant attenuation for the underwater scenario can already be seen for distances greater than 1 m, motivating the use of even lower frequency AC magnetic fields (see inset in Fig.~\hyperref[F4]{\ref{F4}(a)}). The optimal communication frequency will ultimately depend on the tradeoff between the amount of skin depth attenuation affordable and the loss in sensitivity of the NV diamond receiver as the frequency approaches the DC limit and DC magnetometry protocols limited by the spin dephasing time, $T_{2}^{\star}$, rather than $T_{2}$ (i.e., $T_{2}^{\star} \ll T_2)$ must be applied instead. Adhering to Hahn-echo AC magnetometry, an optimum sensitivity is expected to occur around $f_{\mathrm{AC}} {\sim} 1/T_{2}=4.33$ kHz \cite{Taylor2008}. Combining NV magnetometers with magnetic flux concentrators can further enhance the sensitivity by $1-2$ orders of magnitude \cite{Zhu2024,Silani2023}.


At the largest distance considered (d${\sim}0.9$ m), the current ME resonator, which is driven sub-optimally (see Appendix \ref{ME}), is able to produce an AC magnetic field strength of approximately 420 pT at half maximum power (V$_{\mathrm{drive}} = 100$ V). Several improvements can be made to enhance this signal strength, including further geometric and material property optimization and the use of multiple ME resonator elements to form a transmitter array \cite{Dong2025,Chen2024}. In addition, optimization of the signal source and implementation of an impedance matching network to efficiently drive the ME resonators at higher voltages would also yield improvements in magnetic field strength. The projected AC magnetic field strength transmitted through air is plotted as a function of distance for different transmitter array sizes in Fig.~\hyperref[F4]{\ref{F4}(b)}, assuming the current ME resonator design is driven at full power (V$_{\mathrm{drive}} = 200$ V) and the field strength scales approximately linearly with the number of ME resonators within the array. Assuming the NV magnetometer is sampling at $f_\mathrm{samp} = 100$ Hz, the equivalent shot-noise is approximately 3.3 pT (see Appendix~\ref{shot}). With additional improvements to the NV receiver to operate near the shot-noise limit and scaling the current ME resonator array to be greater than 10 elements, magnetic communication over 10 meters in air should be easily achievable.

\section{Conclusion}
In conclusion, we have demonstrated both amplitude and phase-modulated magnetic communication using a hybrid ME resonator transmitter and quantum diamond magnetometer receiver system operating at $f_{\mathrm{AC}} = 20$~kHz. In terms of sensitivity, the amplitude and phase sensitivity of the NV diamond receiver was 50 $\mathrm{pT/\sqrt{Hz}}$ and 1.2 $\mathrm{mrad/\sqrt{Hz}}$ at $B_{\mathrm{AC}}=34.3$ nT, respectively. Finally, we project out and consider communication distances achievable with further optimization of both the transmitter and receiver in relevant conductive mediums.

\begin{acknowledgments}
D. A. B. acknowledges support from the Australian Research Council through grant DE230100192.
\end{acknowledgments}

\newpage

\appendix

\section{ME resonator radiation pattern}
\label{ME_pattern}
The radiation characteristics of the ME resonator design is commonly approximated by an equivalent magnetic moment \cite{Dong2020,Xu2019}. In this approach, at operational frequencies much lower than $\lambda/10$, the ME resonator can be represented by a small circular loop, the plane of which is perpendicular to the magnetization direction of the amorphous metal laminate. Magnetic field contributions from both the respective piezoelectric, $H^E$, and magnetostrictive phases, $H^M$, can be described as \cite{Dong2020,balanis2016}

\begin{equation}
H_{\mathrm{\psi}}^{E}= \frac{p_0k^2}{4\pi}\left(\frac{1}{kr}+\frac{i}{(kr)^2}\right)\sin\theta e^{ikr}
\label{eq:eq2}
\end{equation}
\begin{equation}
H_{r}^{E}= H_{\mathrm{\theta}}^{E}=0
\label{eq:eq2}
\end{equation}
\begin{equation}
H_{r}^{M}= i\frac{m_0k}{2\pi r^2}\left(1+\frac{1}{ikr}\right)\cos\theta e^{-ikr}
\label{eq:eq2}
\end{equation}
\begin{equation}
H_{\mathrm{\theta}}^{M}= \frac{m_0k^2}{2\pi r}\left(1+\frac{1}{ikr}-\frac{1}{(kr)^2}\right)\sin\theta e^{-ikr}
\label{eq:eq2}
\end{equation}
\begin{equation}
H_{\mathrm{\psi}}^{M}=0 \; ,
\label{eq:eq2}
\end{equation}

\noindent{where} $p_0$ and $m_0$ are the total electric and magnetic dipole moments, $k$ is the wave number and $r$, $\theta$ and $\psi$ are the spherical coordinates with respect to the ME resonator geometry shown in Fig.~\hyperref[F1]{\ref{F1}(a)}. The maximum magnetic field generated by the ME resonator is expected to occur when $\theta=0$ such that only $H_{r}^{M}\neq0$. As such, the NV diamond receiver is aligned accordingly with respect to the ME resonator along the z-axis in the laboratory frame (see Fig.~\hyperref[F1]{\ref{F1}(a)}). 


\section{ME resonator characterization}
\label{ME}

\begin{figure}
    \centering
    \includegraphics[width=3.375in]{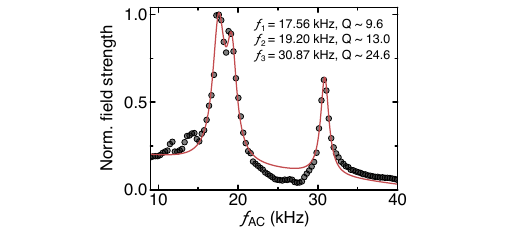}
    \caption{\textbf{ME resonator modes.} Normalized frequency response of the ME resonator. The measured response (black symbols) is fitted using three Lorentzians and a linear polynomial background to determine the resonant frequency and quality factor of each resonance (red solid line).} 
    \label{FS1}
\end{figure}

The normalized frequency response of the ME resonator measured independently using a loop antenna (A.H. Systems, AHS-SAS-560) is presented in Fig.~\hyperref[FS1]{\ref{FS1}}. A complex frequency response is observed, with the ME resonator supporting at least three ME resonances. The normalized frequency response is fitted with three Lorentzians and a linear polynomial background to determine the resonant frequency and quality factor of each resonance (red solid line). Despite the fact that the ME resonator was not operated at one of its resonances for the measurements presented in the main text, our results demonstrate that magnetic communication at $f_{\mathrm{AC}} = 20$~kHz is still achievable in this sub-optimal configuration. Operating at $f_{\mathrm{AC}} = 17.56$ kHz will increase the AC magnetic field strength the ME resonator produces by a factor of ${\sim}2$ compared to operating at$f_{\mathrm{AC}} = 20$ kHz.

\section{AC magnetic field sensitivity} \label{sens}
In the main text, the sensitivity of the NV diamond receiver was evaluated by considering the standard deviation of the histogram distribution of magnetic field values measured, divided by the sampling rate. For Fig.~\hyperref[F3]{\ref{F3}(a)}, this corresponds to $\sigma_{\mathrm{off}}/\sqrt{f_{\mathrm{samp}}}=1.61 \; \mathrm{nT}/\sqrt{1000 \; \mathrm{Hz}}=50.3 \; \mathrm{pT}/\sqrt{\mathrm{Hz}}$. Similarly, the phase sensitivity is calculated from Fig.~\hyperref[F3]{\ref{F3}(b)}, where the measured PL response was converted into phase by taking the inverse sine and using the extracted sinusoidal fit parameters determined from Fig.~\hyperref[F2]{\ref{F2}(d)}.

\begin{figure}
    \centering
    \includegraphics[width=3.375in]{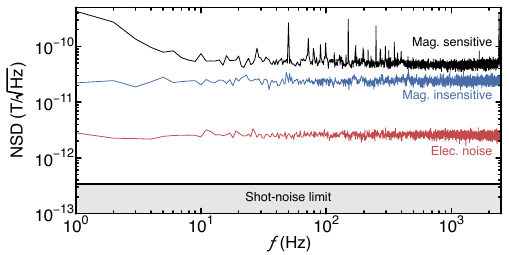}
    \caption{\textbf{Hahn-echo AC magnetometry sensitivity.} Magnetic noise spectral density (NSD) using Hahn-echo AC magnetometery ($\tau=50 \; \upmu$s) calculated for the magnetically sensitive (black), magnetically insensitive (blue), and electronic noise (red) case. The data was measured over 1 s long acquisitions at $f_\mathrm{samp} = 5$ kHz and the NSD for each case was averaged over 30 scans. Horizontal black line indicates shot-noise limit based on the PL measured, accounting for the number of referencing steps implemented (see Appendix \ref{shot} for more details).} 
    \label{FS2}
\end{figure}

An alternative and equivalent method to calculate the sensitivity is by considering the noise spectral density (NSD), which can provide additional frequency content information. This is performed separately by recording 1~s long time traces without the ME resonator at the full sampling rate of $f_\mathrm{samp} = 5$ kHz without decimation and averaging, which are then fast Fourier transformed. To convert the digitized photovoltage into an AC magnetic field strength, the conversion factor of 307 $\upmu$V/nT determined from the inset of Fig.~\hyperref[F2]{\ref{F2}(d)} was used. The averaged NSD (over 30 scans each) within the measurement bandwidth of $f_\mathrm{samp}/2 = 2.5$ kHz is shown in Fig.~\hyperref[FS2]{\ref{FS2}}. 

For the magnetically sensitive case (black solid line) where coherent Hahn-echo control sequence is applied on resonance, a minimum sensitivity of $47.2\pm4.6$~pT/$\sqrt{\mathrm{Hz}}$ is achieved, consistent with the sensitivity value calculated above from the histogram distribution in Fig.~\hyperref[F3]{\ref{F3}(a)}. A pronounced $1/f-$like low frequency noise is observed approximately below 10 Hz. Additionally, discrete frequency components at 50 Hz mains power and its harmonics are observed. The origin of the other discrete frequency components not at multiples of 50 Hz (possibly due to environmental magnetic noise in the laboratory) remains unknown and requires further investigation.

For the magnetically insensitive case (blue solid line), the MW control sequence is detuned off resonance and a minimum sensitivity of $24.3\pm2.7$~pT/$\sqrt{\mathrm{Hz}}$ is achieved. Furthermore, the $1/f-$like low frequency noise and the discrete frequency components have disappeared, indicating their origin is magnetic in nature. The electronic noise (red solid line) from the photodetectors and analog-to-digital converter is characterized with the laser turned off and a sensitivity of $2.6\pm0.3$~pT/$\sqrt{\mathrm{Hz}}$ is achieved. The shot-noise limited sensitivity is independently calculated to be approximately $0.33$~pT/$\sqrt{\mathrm{Hz}}$ based on the total amount of PL collected (see Appendix \ref{shot}). This indicates significant residual PL noise (which include both laser intensity noise and NV charge state fluctuations) still remains despite implementing two photodetectors for common-mode rejection. Additionally, the current electronics front end and its limited dynamic range will prevent operation near the shot-noise limit even in the ideal case where laser intensity noise is fully suppressed. An integrating balanced photodetector implemented in Ref.~\cite{Barry2024} particularly for pulsed operation will likely enable operation near the shot-noise limit.

\section{Shot-noise limited sensitivity} \label{shot}
The photon shot-noise level based on the measured PL from the NV diamond is calculated as follows. Accounting for the reduced signal measured during pulsed laser readout (i.e., $\Delta t/T_{seq} = 30 \; \upmu\mathrm{s}/100 \; \upmu\mathrm{s}$), the voltage shot-noise is \cite{Wolf2015}:

\begin{equation}
\sigma_{\mathrm{shot}} = G \sqrt{2 q I_{PL}  f \cdot (\Delta t/T_{seq})} \; ,
\label{eq:eq2}
\end{equation}

\noindent{where} $G = 1510$ V/A is the photodetector gain, $q$ is the electronic charge, $I_{PL} = 11.9 \;\mathrm{mW} \cdot 0.5 \; \mathrm{A/W} = 5.85$~mA the measured photocurrent, and $f$ is the measurement bandwidth. Inserting the values into Eq.~\ref{eq:eq2}, the voltage shot-noise per square root unit of bandwidth is 36.1~nV/$\sqrt{\mathrm{Hz}}$. The photovoltage from the reference photodetector is expected to contribute to the overall shot-noise as a separate uncorrelated noise source. In addition, the 180$^{\circ}$ phase cycling referencing step implemented on the final $\pi/2$ pulse introduces another uncorrelated noise term. Accounting for the two referencing steps described above and an additional factor for doubling the measurement time \cite{Wolf2015}, shot-noise is expected to increase by a factor of $\sqrt{2^{3}}$ due to the overall contribution from all uncorrelated noise sources in the system. Therefore, the shot-noise limited sensitivity equates to $\big(\sqrt{2^{3}} \cdot 36.1 \; \mathrm{nV}/\sqrt{\mathrm{Hz}} \big)/ 307 \; \mathrm{\upmu V/nT} = 0.33 \; \mathrm{pT}/\sqrt{\mathrm{Hz}}$.

\newpage
\bibliography{apssamp}

\end{document}